\documentclass[
]{ceurart}
\graphicspath{ {./images/} }

\begin{document}
\copyrightyear{2021}
\copyrightclause{Copyright for this paper by its authors.
  Use permitted under Creative Commons License Attribution 4.0
  International (CC BY 4.0).}

\conference{Perspectives on the Evaluation of Recommender Systems Workshop (PERSPECTIVES 2021), September 25th, 2021,
co-located with the 15th ACM Conference on Recommender Systems, Amsterdam, The Netherlands}

\title{Statistical Inference: The Missing Piece of RecSys Experiment Reliability Discourse}

\author[1]{Ngozi Ihemelandu}[%
email=ngoziihemelandu@u.boisestate.edu,
]
\address[1]{People and Information Research Team, Boise State University,
  1910 University Drive, Boise, Idaho, USA, 83725-2055}

\author[2]{Michael D. Ekstrand}[%
orcid=0000-0003-2467-0108,
email=michaelekstrand@boisestate.edu,
url=https://md.ekstrandom.net/,
]
\address[2]{People and Information Research Team, Boise State University,
  1910 University Drive, Boise, Idaho, USA, 83725-2055}



\begin{abstract}
This paper calls attention to the missing component of the recommender system evaluation process: Statistical Inference. There is active research in several components of the recommender system evaluation process: selecting baselines, standardizing benchmarks, and target item sampling. However, there has not yet been significant work on the role and use of statistical inference for analyzing recommender system evaluation results.

In this paper, we argue that the use of statistical inference is a key component of the evaluation process that has not been given sufficient attention. We support this argument with systematic review of recent RecSys papers to understand how statistical inference is currently being used, along with a brief survey of studies that have been done on the use of statistical inference in the information retrieval community.
We present several challenges that exist for inference in recommendation experiment which buttresses the need for empirical studies to aid with appropriately selecting and applying statistical inference techniques.

\end{abstract}






\maketitle

\section{Introduction}
It is widely recognized that the use of appropriate statistical inference techniques should be used to analyze, interpret, and report the results of evaluations and experiments, including evaluations of recommender systems \citep{konstan2013toward}. These techniques come in many forms, including point estimation, interval estimation, and hypothesis testing, but analysis needs to go beyond merely computing metrics to determine if observed metrics represent genuine effects.  \footnote{The video accompanying this paper is available at \url{https://piret.info/pubs/2021/Perspectives21-StatisticalInference}} In this paper we consider the state of statistical inference in recommender systems evaluation, arguing that identifying and documenting best practices for statistical analysis is a vital and oft-overlooked component of the discussion on how to improve the rigor, reproducibility, and reliability of recommender systems evaluation results.

We focus primarily on statistical inference for one of the most common goals of recommender systems research: to demonstrate an improvement in effectiveness over the current state of the art. This could be by developing a new recommendation technique that is more effective at some recommendation tasks than previously-known techniques, or by modifying an existing approach. To assess if the measured improvement of the new method over the state-of-the-art is substantial and not just a result of random chance, we typically use a hypothesis test (null hypothesis significance testing, or NHST) for the null hypothesis that there is no difference between the two methods' effectiveness; sometimes confidence intervals or Bayesian inference techniques may be employed instead of or in addition to an NHST.

There has been significant research on evaluation strategies for this research goal.
\citet{ dacrema2019we} showed in their systematic analysis of deep learning approaches for top-$N$ recommendation tasks that many claims of improved performance over a baseline may be illusory.
There are many design points in a recommender experiment that can affect its rigor and reliability; \citeauthor{dacrema2019we} focused specifically on the choice and tuning of baselines in the evaluation process. They found that many measured improvements disappear when the baseline algorithms are properly tuned: that is, better choice of hyperparameters and model options can cause the baseline to perform just as well as the proposed new method. 

Other authors have considered the effects and sought to develop best practices for other design choices in an evaluation. \citet{rendle2019difficulty} argue for standardized benchmarks, by which they mean datasets with well-defined train–test splits and evaluation protocols for specific tasks (e.g. prediction). They state that although well-defined benchmarks exist for comparing prediction algorithms, there are not standardized benchmarks for other tasks such as ranking. They argue that empirical findings reported in research papers are questionable unless they were obtained on standardized benchmarks where --- as recommended by \citeauthor{dacrema2019we} --- baselines have been tuned extensively. \citet{canamares2020target} bring attention to an offline evaluation setup component --- target item sampling --- that is not always explicit and has received little attention in the quest for seeking an evaluation procedure. They show that different target subsets can lead to different evaluation outcomes. \citet{sun2020we} work shed light on the issues -- unreproducible evaluation and unfair comparison -- which they attribute to the unavailability of effective benchmarks for evaluation. They investigated the evaluation rigorousness (reproducibility and fairness) in recommendation by analyzing the influence of different factors on recommendation performance through a holistic empirical study. The result of their study corroborates the findings of \citet{dacrema2019we}.

However, there has not yet been much attention to appropriately selecting and applying statistical inference techniques to the metrics that result from these evaluations. \citet{shani2011evaluating} discuss general ways of performing significance testing using widely-known statistical methods, but to our knowledge there have not yet been empirical studies on the use of statistical inference for analyzing evaluation results, as there has been for TREC-style search experiments (see Section~\ref{sec:ir-studies}).  Evidence-based guidance on best practices for analyzing and reporting results is therefore lacking.  The current use, or lack thereof, of various techniques for recommender system experimental results is also an open question.

Our central claim in this paper is that the RecSys community does not currently pay sufficient attention to the choice and use of statistical techniques, and discussions such as the one at this workshop needs to consider the role of inference and develop best practices for rigorous analysis of evaluation results. We support this argument with a systematic review of recent RecSys papers to understand how statistical inference is currently being used, along with a brief survey of studies that have been done on the use of statistical inference in the information retrieval community (particularly for analyzing TREC search effectiveness metrics). We identify several challenges that exist for inference in recommendation experiments, and call on the community to attend to this issue and work with us to fill this important gap in the literature on reliable evaluation of recommender systems.

\section{Systematic Review of Statistical Inference in RecSys}
\label{sec:review}

We begin by assessing current practices in statistical inference for recommender system evaluations. Our study is inspired by that of \citet{sakai2016statistical}, who conducted a systematic review of 840 SIGIR full papers and 215 TOIS papers published between 2006 and 2015. Their goal was to identify what types of statistical test IR researchers use, how they report or fail to report on significance test results, and how the reporting practices may have changed over the last decade.

They found that of the 862 papers selected for the survey about 28-30\% do not report significance test results; for the comparison of two IR systems, 61-66\% of these papers use the paired $t$-test; 20-23\% use the Wilcoxon signed rank test; 4-5\% use the randomisation test; 3-4\% use the sign test; and 1\% use the bootstrap test. They also found that the paired $t$-test was more common in recent years while Wilcoxon test decreased in popularity.

To get a first look at current RecSys statistical practices, we conducted a systematic review only for long and short RecSys papers that proposed new or enhanced algorithmic methods and compared their performance to that of baselines (state-of-art). Hence, proposed new methods that were not compared to baselines were not selected. Our survey is limited to papers published in 2019 and 2020. 

\subsection{Survey Methods}
The main focus of this systematic survey is to examine how statistical significance tests are used by researchers working on papers proposing new or enhanced recommender algorithms.

We selected full and short papers from RecSys 2019--2020 that meet the following criteria:
\begin{itemize}
    \item The paper proposed a new or enhanced algorithmic method for some recommendation task.
    \item The effectiveness scores for the baselines and new/enhanced method were recorded.
\end{itemize}

We coded the selected papers as specified below (The coding was done in the listed order. That is, if the paper does not meet the first criteria, the second criteria is checked etc.):
\begin{description}
    \item[Used specified test]
    The paper mentioned the name of the test used along with the significance level ($\alpha$) or $p$-value.  We also recorded which test it used.
    
    \item[Used confidence interval]
    The paper reported confidence intervals or indicated the standard error for the estimated metric scores for the new method as well as the baseline.
    
    \item[Used unspecified test]
    The paper did not specify which test was used but claimed statistical significance or specified  $p$-value $<$ significance level ($\alpha$) or the calculated test statistics.
    
    \item[No significance test]
    The paper did not seem to test the results for significance.
\end{description}

\subsection{Results and Discussion}

Out of the 146 RecSys long and short papers examined, we found 111 papers that proposed new or enhanced recommender algorithms for which we expect significance testing to be used to analyze the evaluation result. See Table~\ref{tbl:year-papers} for the break down of the selected papers by year.

\begin{table}[tb]
    \caption{Long and short algorithm papers by year}
    \label{tbl:year-papers}
    \centering
    \begin{tabular}{cccc}
        Year&\# long papers&\# short papers& long + short\\
        \toprule
        2020 & 25 & 21 & 46 \\
        2019 & 36 & 29 & 65\\
        \textsf{Total} &  & & 111\\
   \end{tabular}
\end{table}

\begin{table}[tb]
    \caption{Paper count and percentage for each category}
    \label{tbl:test-status}
    \centering
    \begin{tabular}{ccc}
        Category&\# Papers&\% of Papers\\
        \toprule
        \textsf{Used specified test} & 26 & 23  \\
        \textsf{Used confidence interval} & 8 & 7  \\
        \textsf{Used unspecified test} & 12 & 11  \\
        \textsf{No significance test} & 65 & 59 \\
   \end{tabular}
\end{table}

\begin{figure}[tb]
\centering
\includegraphics[scale=0.5]{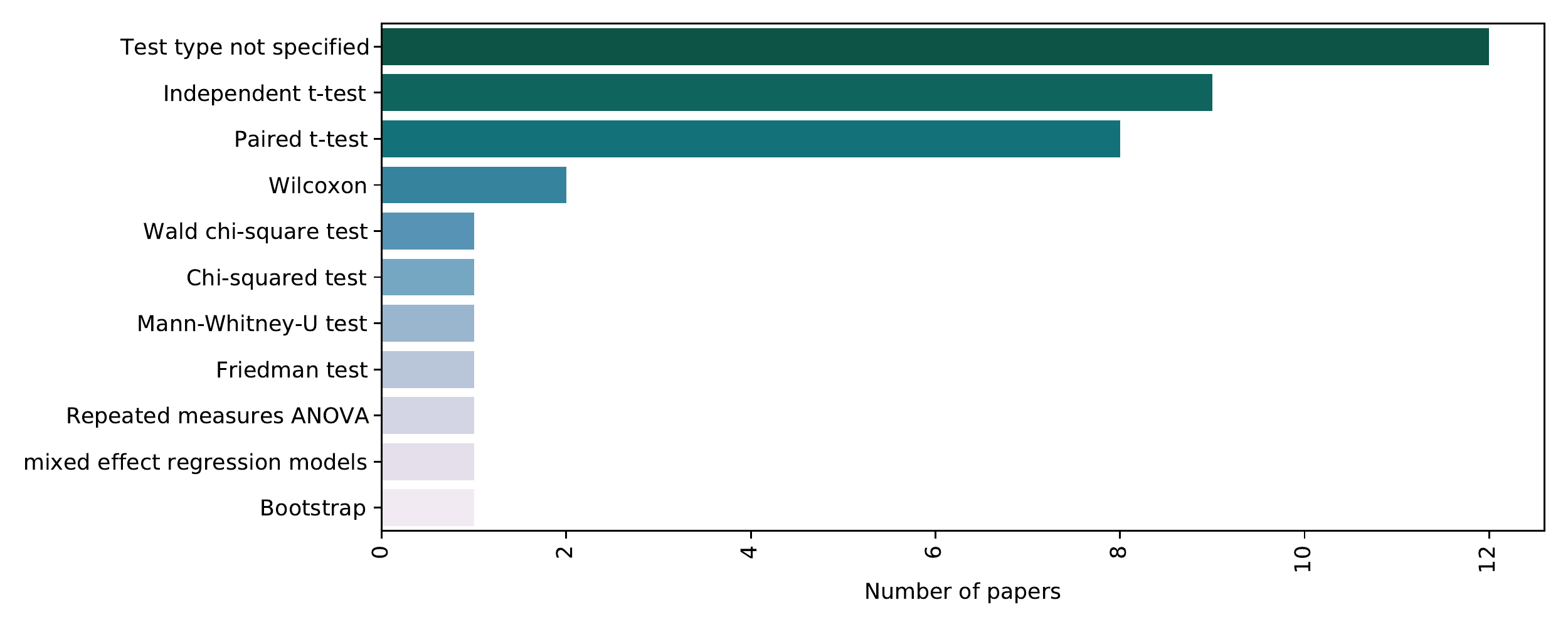}
\caption{Distribution of selected papers coded as -- used significance test.}
\label{fig:testType}
\end{figure}

Table \ref{tbl:test-status} shows the classification of the 111 selected papers, and Fig.~\ref{fig:testType} shows the distribution by the test type of the set of selected papers labeled as ``used significance tests''.
We found that over half of the papers proposing a new algorithmic method did not seem to use any significance test to analyze their evaluation results; a substantial portion of those who claim significance did not specify a test.

These results show that there is currently a lack of rigorous statistical analysis and reporting in the evaluations published in RecSys.  While we do not have an explanation as to why there is this gap, we believe it needs to be filled if we are to go from observed differences in metrics to reliable knowledge.

\section{Statistical Inference in Information Retrieval}
\label{sec:ir-studies}

In addition to \citeauthor{sakai2016statistical}'s study of existing practice, several studies in the information retrieval (IR) community have addressed the use of statistical inference for system comparison experiment, attempting to identify which statistical techniques are appropriate to use for the analysis of the evaluation results in IR systems comparison, particularly for the results of TREC-style experiments.

\citet{smucker2007comparison} used results from historical TREC runs to study the agreement between different pairwise significance tests.  Using root mean squared error (RMSE) to compare the $p$-values produced by five different tests, they found that the randomization, bootstrap, and $t$-tests all agreed with each other (producing very similar $p$-values) while the Wilcoxon and sign tests neither agreed with the other tests nor each other.
They then used the randomization test as ground truth to estimate the false positive and false negative rates of the Wilcoxon and Sign tests, finding that both tests have high false positive and false negative rates when the difference in system effectiveness (evaluation metric) is small. 
They recommend that researchers wanting a distribution-free test should use the randomization test with the test statistic of their
choice, and recommended discontinuing use of the Wilcoxon or sign tests for IR evaluation data analysis.

\citet{urbano2019statistical} and \citet{parapar2020using} used simulations to produce per-topic evaluation scores rather than directly using the recorded metrics. They fit generative probabilistic models to the metric distributions from historic TREC runs (to ensure realism) and sampled from these models, allowing them to directly control the actual difference (or lack thereof) between systems and measure the error rates of different statistical tests. One of the key differences in their approaches is the simulation architecture: \citet{parapar2020using} simulated the utility of individual retrieved documents, while \citet{urbano2019statistical} modeled the joint distribution between pairs of effectiveness scores.
Both simulation designs enabled them to directly assess the accuracy of the p-values produced by the various significance tests, and to measure their false positive rates and statistical power.

\citet{urbano2019statistical} found that the Wilcoxon and sign tests have more false positives than expected, especially at low significance levels, and that this error is more pronounced as the sample size increases. The bootstrap test exhibits similar behavior (making more false positive rates than expected) with small sample sizes but starts behaving as expected as the sample size increases. The randomization test behaves better than the bootstrap, Wilcoxon and sign tests and approaches the expected behavior as the sample size increases. The $t$-test behaves as expected even for small sample size. They also found that for large sample sizes the randomization, bootstrap and $t$-test all agree, concurring with the results of \citet{smucker2007comparison}.

They also found that the sign test is consistently less powerful than other tests while the bootstrap test is usually the most powerful, especially with small samples. With large sample sizes, all tests except the sign tests exhibited nearly-identical power. Since the $t$-test was well-behaved as  in terms of both the false positive rate (even for small samples) and power, \citeauthor{urbano2019statistical} recommend its use as the best choice for mean effectiveness in IR evaluations, and the randomization test for test statistics other than the mean. Like \citeauthor{smucker2007comparison}, they discourage use of the Wilcoxon and sign tests for IR evaluation results.

\citet{parapar2020using} came to different conclusions than \citeauthor{urbano2019statistical}. Their simulations showed that the Wilcoxon and randomization tests have the expected false-positive rate behavior while the $t$-test, the sign test, and the bootstrap did not behave as expected. They also found that the sign test and Wilcoxon test have more statistical power than the other tests. Therefore they recommend the use of the sign test and Wilcoxon test for the analysis of IR evaluation results.

All three papers had the goal of producing recommendations for appropriate significance tests to apply when comparing IR systems.
\citet{smucker2007comparison} and \citet{urbano2019statistical} made recommendations that were similar, while \citet{parapar2020using} arrived at a completely different recommendation. 

Both \citet{urbano2021metric}  and \citet{parapar2021testing} have followed up and attempted to understand this discrepancy in their conclusions, but there is not yet clarity on which is the more reliable recommendation.

\section{Gaps for Fixing RecSys Evaluation Practice}

Whichever evidence produces the more reliable recommendation for IR evaluation settings studied in the previous section, it may not be feasible to just apply that recommendation to RecSys evaluation. There are some key differences between TREC ad-hoc retrieval evaluation and the recommender system evaluation. Some of these key differences --- which do not only affect recommendation, as many are shared with actual deployments of search engines outside the TREC context --- include:

\begin{itemize}
    \item The sample size of the test collection in a traditional TREC Cranfield experiment is quite small --- often 50 topics, particularly in the data sets studied --- while the typical sample size of a RecSys evaluation is $> 1,000$ users.
    \item In typical RecSys evaluation data, a few items are known to be relevant to many users, resulting in a long-tailed distribution of user ratings over items. This is in contrast with TREC evaluation where documents are not concentrated to just a few queries.
    \item In TREC evaluation, the ground-truth relevance judgement which are assumed to be (approximately) complete, while the user feedback used in RecSys evaluations form a sparse and highly incomplete picture of item-user relevance.
\end{itemize}

We want to call particular attention to sample size, as it is a key factor that impacts the statistical power of a significance test (the ability of the test to detect significance in the presence of a real effect). The statistical power of a significance test increases as the sample size increases; therefore, by increasing the sample size, any measured improvement can be found to be significant by any significance test even when the size of the measured improvement is so small that it is not operationally meaningful.

Statistical biases are another factor that may influence the outcome of significance test for RecSys evaluation data. It has become well known that biases such as sparsity and popularity biases in RecSys evaluation data considerably distort the evaluation measures \citep{ekstrand2017sturgeon,tian2020estimating, canamares2017probabilistic, canamares2018should, canamares2020target}. \citet{bellogin2017statistical} showed that the long-tailed distribution of RecSys evaluation data has a drastic effect on how recommendation algorithms compare to each other.
The hypothesis test does not account for these biases hence, this distortion can ultimately influence its outcome.  It isn't clear whether this should be fixed as a part of inference, or as a corrective stage before or after inference, but it remains a gap in the ability to accurately evaluate system performance that needs to be addressed.

There are also on-going discussions on the inadequacies of statistical significance testing. \citet{mcshane2019abandon} states that the widespread crisis in the biomedical and social sciences with published findings failing to replicate at an alarming rate maybe associated with claims of huge effects from tiny interventions, citing $p < 0.05$ as the primary evidence.
A group of 72 researchers representing a wide range of disciplines (psychology, economics, sociology, anthropology, medicine, epidemiology, ecology, and philosophy) and statistical perspectives have proposed a change in the $p$-value threshold for a “statistically significant” result from 0.05 to 0.005 for claims of discoveries of novel effects \citep{benjamin2018redefine}. They recommend that results currently called “statistically significant” that do not meet the new threshold would be called suggestive and treated as ambiguous as to whether there is an effect. However, \citet{mcshane2019abandon} state that this proposal is insufficient to overcome the current crisis with the inability to replicate experiment results. They recommend abandoning the null hypothesis significance testing paradigm entirely and just use $p$-values as one of many pieces of information to cite as evidence for a novel effect claim.

Translating this discussion back to information retrieval, \citet{sakai2014statistical} recognizes that statistical significance testing is not enough and provides suggestions on how IR researchers should report effect sizes and confidence intervals along with $p$-values, in the context of comparing IR systems using test collections. 
\citet{ carterette2012multiple} advocates for the use of the $t$-test even though their analysis showed that a $p$-value cannot have any objective meaning. They believe it is still useful for many of the purposes they are currently used for. They however, recommend that in the long term, IR experimental analysis should transition to a fully Bayesian modeling approach.

We raise these points to observe that even if we can identify effective and appropriate hypothesis tests for typical offline evaluation metrics, that does not fully address the goal of inferring whether or not a proposed system is actually more effective; additional sources of bias need to be accounted for, and it is not clear that NHST is the best framework for evaluating results.

\section{Challenges of Statistical Inference and Next Steps}

It is important that statistical inference results are reported with all necessary details in order to make research papers as informative as possible, and to estimate and give the reader confidence in understanding the credibility and impact of a reported improvement. However, this is not the prevalent current practice in the RecSys community, as demonstrated by the results in Section~\ref{sec:review}. While there has been significant attention paid to other aspects of the evaluation process 
\cite{dacrema2019we, rendle2019difficulty, canamares2020target, sun2020we}, and the IR community has studied inference for certain experimental settings (see Section~\ref{sec:ir-studies}), this aspect has not yet been a noticeable part of the scholarly discourse on evaluation practices.  We argue that this gap needs to be filled.

As a first step, we propose that researchers should report clearly how they performed inference on their results, with multiple results as appropriate. For example, studies using frequentist significance testing should report the test used, the $p$-value threshold, any corrections for multiple comparisions, and also the effect size and confidence interval, in order to help readers fully understand and better apply the findings. Reporting effect size and sample size help to make papers as informative as possible. While further research is needed to identify best practices for selecting and applying techniques, research using current practices should clearly document them.

We believe further research is needed to identify best practices for applying and reporting on classical tests and techniques, and to study how more advanced inference techniques may be able to mitigate some of their limitations. One such advanced technique that could be studied in the recommender system context is the mixed effect model for testing significance of effects, or Bayesian inference techniques for computing and summarizing posterior distributions of effect sizes.
We also believe that, as the community continues work towards documented best practices, and has discussed in the past the need to lay out recommended methods for the benefit of authors, reviewers, and editors \citep{konstan2013toward}, such practices need to include recommendations for statistical techniques.
The community may be ready to make some such recommendations now, but we call for further research to provide empirical evidence for the appropriateness of recommended techniques, and for such guidelines to leave the door open for innovation in statistical analysis of recommender system evaluations, at least so long as the direction of this innovation is towards greater understanding and rigor. 

\begin{acknowledgments}
This work partially supported by the National Science Foundation under Grant IIS 17-51278.
\end{acknowledgments}


\bibliography{reference}

\begin{thebibliography}{21}
\expandafter\ifx\csname natexlab\endcsname\relax\def\natexlab#1{#1}\fi
\providecommand{\url}[1]{\texttt{#1}}
\providecommand{\href}[2]{#2}
\providecommand{\path}[1]{#1}
\providecommand{\DOIprefix}{doi:}
\providecommand{\ArXivprefix}{arXiv:}
\providecommand{\URLprefix}{URL: }
\providecommand{\Pubmedprefix}{pmid:}
\providecommand{\doi}[1]{\href{http://dx.doi.org/#1}{\path{#1}}}
\providecommand{\Pubmed}[1]{\href{pmid:#1}{\path{#1}}}
\providecommand{\bibinfo}[2]{#2}
\ifx\xfnm\relax \def\xfnm[#1]{\unskip,\space#1}\fi
\bibitem[{Konstan and Adomavicius(2013)}]{konstan2013toward}
\bibinfo{author}{J.~A. Konstan}, \bibinfo{author}{G.~Adomavicius},
\newblock \bibinfo{title}{Toward identification and adoption of best practices
  in algorithmic recommender systems research},
\newblock in: \bibinfo{booktitle}{Proceedings of the international workshop on
  Reproducibility and replication in recommender systems evaluation},
  \bibinfo{year}{2013}, pp. \bibinfo{pages}{23--28}.
\bibitem[{Dacrema et~al.(2019)Dacrema, Cremonesi, and Jannach}]{dacrema2019we}
\bibinfo{author}{M.~F. Dacrema}, \bibinfo{author}{P.~Cremonesi},
  \bibinfo{author}{D.~Jannach},
\newblock \bibinfo{title}{Are we really making much progress? a worrying
  analysis of recent neural recommendation approaches},
\newblock in: \bibinfo{booktitle}{Proceedings of the 13th ACM Conference on
  Recommender Systems}, \bibinfo{year}{2019}, pp. \bibinfo{pages}{101--109}.
\bibitem[{Rendle et~al.(2019)Rendle, Zhang, and Koren}]{rendle2019difficulty}
\bibinfo{author}{S.~Rendle}, \bibinfo{author}{L.~Zhang},
  \bibinfo{author}{Y.~Koren},
\newblock \bibinfo{title}{On the difficulty of evaluating baselines: A study on
  recommender systems},
\newblock \bibinfo{journal}{arXiv preprint arXiv:1905.01395}
  (\bibinfo{year}{2019}).
\bibitem[{Ca{\~n}amares and Castells(2020)}]{canamares2020target}
\bibinfo{author}{R.~Ca{\~n}amares}, \bibinfo{author}{P.~Castells},
\newblock \bibinfo{title}{On target item sampling in offline recommender system
  evaluation},
\newblock in: \bibinfo{booktitle}{Fourteenth ACM Conference on Recommender
  Systems}, \bibinfo{year}{2020}, pp. \bibinfo{pages}{259--268}.
\bibitem[{Sun et~al.(2020)Sun, Yu, Fang, Yang, Qu, Zhang, and Geng}]{sun2020we}
\bibinfo{author}{Z.~Sun}, \bibinfo{author}{D.~Yu}, \bibinfo{author}{H.~Fang},
  \bibinfo{author}{J.~Yang}, \bibinfo{author}{X.~Qu},
  \bibinfo{author}{J.~Zhang}, \bibinfo{author}{C.~Geng},
\newblock \bibinfo{title}{Are we evaluating rigorously? benchmarking
  recommendation for reproducible evaluation and fair comparison},
\newblock in: \bibinfo{booktitle}{Fourteenth ACM Conference on Recommender
  Systems}, \bibinfo{year}{2020}, pp. \bibinfo{pages}{23--32}.
\bibitem[{Shani and Gunawardana(2011)}]{shani2011evaluating}
\bibinfo{author}{G.~Shani}, \bibinfo{author}{A.~Gunawardana},
\newblock \bibinfo{title}{Evaluating recommendation systems},
\newblock in: \bibinfo{booktitle}{Recommender systems handbook},
  \bibinfo{publisher}{Springer}, \bibinfo{year}{2011}, pp.
  \bibinfo{pages}{257--297}.
\bibitem[{Sakai(2016)}]{sakai2016statistical}
\bibinfo{author}{T.~Sakai},
\newblock \bibinfo{title}{Statistical significance, power, and sample sizes: A
  systematic review of sigir and tois, 2006-2015},
\newblock in: \bibinfo{booktitle}{Proceedings of the 39th International ACM
  SIGIR conference on Research and Development in Information Retrieval},
  \bibinfo{year}{2016}, pp. \bibinfo{pages}{5--14}.
\bibitem[{Smucker et~al.(2007)Smucker, Allan, and
  Carterette}]{smucker2007comparison}
\bibinfo{author}{M.~D. Smucker}, \bibinfo{author}{J.~Allan},
  \bibinfo{author}{B.~Carterette},
\newblock \bibinfo{title}{A comparison of statistical significance tests for
  information retrieval evaluation},
\newblock in: \bibinfo{booktitle}{Proceedings of the sixteenth ACM conference
  on Conference on information and knowledge management}, \bibinfo{year}{2007},
  pp. \bibinfo{pages}{623--632}.
\bibitem[{Urbano et~al.(2019)Urbano, Lima, and
  Hanjalic}]{urbano2019statistical}
\bibinfo{author}{J.~Urbano}, \bibinfo{author}{H.~Lima},
  \bibinfo{author}{A.~Hanjalic},
\newblock \bibinfo{title}{Statistical significance testing in information
  retrieval: an empirical analysis of type i, type ii and type iii errors},
\newblock in: \bibinfo{booktitle}{Proceedings of the 42nd International ACM
  SIGIR Conference on Research and Development in Information Retrieval},
  \bibinfo{year}{2019}, pp. \bibinfo{pages}{505--514}.
\bibitem[{Parapar et~al.(2020)Parapar, Losada, Presedo-Quindimil, and
  Barreiro}]{parapar2020using}
\bibinfo{author}{J.~Parapar}, \bibinfo{author}{D.~E. Losada},
  \bibinfo{author}{M.~A. Presedo-Quindimil}, \bibinfo{author}{A.~Barreiro},
\newblock \bibinfo{title}{Using score distributions to compare statistical
  significance tests for information retrieval evaluation},
\newblock \bibinfo{journal}{Journal of the Association for Information Science
  and Technology} \bibinfo{volume}{71} (\bibinfo{year}{2020})
  \bibinfo{pages}{98--113}.
\bibitem[{Urbano et~al.(2021)Urbano, Corsi, and Hanjalic}]{urbano2021metric}
\bibinfo{author}{J.~Urbano}, \bibinfo{author}{M.~Corsi},
  \bibinfo{author}{A.~Hanjalic},
\newblock \bibinfo{title}{How do metric score distributions affect the type i
  error rate of statistical significance tests in information retrieval?},
\newblock in: \bibinfo{booktitle}{Conference on the Theory of Information
  Retrieval (ICTIR’21)}, \bibinfo{year}{2021}.
\bibitem[{Parapar et~al.(2021)Parapar, Losada, and
  Barreiro}]{parapar2021testing}
\bibinfo{author}{J.~Parapar}, \bibinfo{author}{D.~E. Losada},
  \bibinfo{author}{{\'A}.~Barreiro},
\newblock \bibinfo{title}{Testing the tests: simulation of rankings to compare
  statistical significance tests in information retrieval evaluation},
\newblock in: \bibinfo{booktitle}{Proceedings of the 36th Annual ACM Symposium
  on Applied Computing}, \bibinfo{year}{2021}, pp. \bibinfo{pages}{655--664}.
\bibitem[{Ekstrand and Mahant(2017)}]{ekstrand2017sturgeon}
\bibinfo{author}{M.~D. Ekstrand}, \bibinfo{author}{V.~Mahant},
\newblock \bibinfo{title}{Sturgeon and the cool kids: Problems with random
  decoys for top-n recommender evaluation},
\newblock in: \bibinfo{booktitle}{The Thirtieth International Flairs
  Conference}, \bibinfo{year}{2017}.
\bibitem[{Tian and Ekstrand(2020)}]{tian2020estimating}
\bibinfo{author}{M.~Tian}, \bibinfo{author}{M.~D. Ekstrand},
\newblock \bibinfo{title}{Estimating error and bias in offline evaluation
  results},
\newblock in: \bibinfo{booktitle}{Proceedings of the 2020 Conference on Human
  Information Interaction and Retrieval}, \bibinfo{year}{2020}, pp.
  \bibinfo{pages}{392--396}.
\bibitem[{Ca{\~n}amares and Castells(2017)}]{canamares2017probabilistic}
\bibinfo{author}{R.~Ca{\~n}amares}, \bibinfo{author}{P.~Castells},
\newblock \bibinfo{title}{A probabilistic reformulation of memory-based
  collaborative filtering: Implications on popularity biases},
\newblock in: \bibinfo{booktitle}{Proceedings of the 40th International ACM
  SIGIR Conference on Research and Development in Information Retrieval},
  \bibinfo{year}{2017}, pp. \bibinfo{pages}{215--224}.
\bibitem[{Ca{\~n}amares and Castells(2018)}]{canamares2018should}
\bibinfo{author}{R.~Ca{\~n}amares}, \bibinfo{author}{P.~Castells},
\newblock \bibinfo{title}{Should i follow the crowd? a probabilistic analysis
  of the effectiveness of popularity in recommender systems},
\newblock in: \bibinfo{booktitle}{The 41st International ACM SIGIR Conference
  on Research \& Development in Information Retrieval}, \bibinfo{year}{2018},
  pp. \bibinfo{pages}{415--424}.
\bibitem[{Bellog{\'\i}n et~al.(2017)Bellog{\'\i}n, Castells, and
  Cantador}]{bellogin2017statistical}
\bibinfo{author}{A.~Bellog{\'\i}n}, \bibinfo{author}{P.~Castells},
  \bibinfo{author}{I.~Cantador},
\newblock \bibinfo{title}{Statistical biases in information retrieval metrics
  for recommender systems},
\newblock \bibinfo{journal}{Information Retrieval Journal} \bibinfo{volume}{20}
  (\bibinfo{year}{2017}) \bibinfo{pages}{606--634}.
\bibitem[{McShane et~al.(2019)McShane, Gal, Gelman, Robert, and
  Tackett}]{mcshane2019abandon}
\bibinfo{author}{B.~B. McShane}, \bibinfo{author}{D.~Gal},
  \bibinfo{author}{A.~Gelman}, \bibinfo{author}{C.~Robert},
  \bibinfo{author}{J.~L. Tackett},
\newblock \bibinfo{title}{Abandon statistical significance},
\newblock \bibinfo{journal}{The American Statistician} \bibinfo{volume}{73}
  (\bibinfo{year}{2019}) \bibinfo{pages}{235--245}.
\bibitem[{Benjamin et~al.(2018)Benjamin, Berger, Johannesson, Nosek,
  Wagenmakers, Berk, Bollen, Brembs, Brown, Camerer
  et~al.}]{benjamin2018redefine}
\bibinfo{author}{D.~J. Benjamin}, \bibinfo{author}{J.~O. Berger},
  \bibinfo{author}{M.~Johannesson}, \bibinfo{author}{B.~A. Nosek},
  \bibinfo{author}{E.-J. Wagenmakers}, \bibinfo{author}{R.~Berk},
  \bibinfo{author}{K.~A. Bollen}, \bibinfo{author}{B.~Brembs},
  \bibinfo{author}{L.~Brown}, \bibinfo{author}{C.~Camerer}, et~al.,
\newblock \bibinfo{title}{Redefine statistical significance},
\newblock \bibinfo{journal}{Nature human behaviour} \bibinfo{volume}{2}
  (\bibinfo{year}{2018}) \bibinfo{pages}{6--10}.
\bibitem[{Sakai(2014)}]{sakai2014statistical}
\bibinfo{author}{T.~Sakai},
\newblock \bibinfo{title}{Statistical reform in information retrieval?},
\newblock in: \bibinfo{booktitle}{ACM SIGIR Forum},
  volume~\bibinfo{volume}{48}, \bibinfo{organization}{ACM New York, NY, USA},
  \bibinfo{year}{2014}, pp. \bibinfo{pages}{3--12}.
\bibitem[{Carterette(2012)}]{carterette2012multiple}
\bibinfo{author}{B.~A. Carterette},
\newblock \bibinfo{title}{Multiple testing in statistical analysis of
  systems-based information retrieval experiments},
\newblock \bibinfo{journal}{ACM Transactions on Information Systems (TOIS)}
  \bibinfo{volume}{30} (\bibinfo{year}{2012}) \bibinfo{pages}{1--34}.

\end{thebibliography}
\end{document}